\documentstyle[12pt,epsfig]{article}
\begin{document}

\begin{flushright}
\framebox{\bf hep-ph/0201151} \\
{BIHEP--TH--2002--5}
\end{flushright}

\vspace{0.2cm}

\begin{center}
{\large\bf Texture Zeros and Majorana Phases of the Neutrino Mass Matrix}
\end{center}

\vspace{0.2cm}

\begin{center}
{\bf Zhi-zhong Xing} \footnote{Electronic address:
xingzz@mail.ihep.ac.cn} \\
{\it Institute of High Energy Physics, P.O. Box 918, Beijing 100039,
China}
\end{center}

\vspace{2cm}

\begin{abstract}
We present the generic formulas to calculate the ratios of neutrino
masses and the Majorana phases of CP violation from the neutrino
mass matrix with two independent vanishing entries in the flavor basis
where the charged lepton mass matrix is diagonal. An order-of-magnitude
illustration is given for seven experimentally acceptable textures
of the neutrino mass matrix, and some analytical approximations are made
for their phenomenological consequences at low energy scales.
\end{abstract}

\newpage

\framebox{\Large\bf 1} ~
The atmospheric and solar neutrino oscillations observed
in the Super-Kamiokande experiment \cite{SK} have
provided robust evidence that neutrinos are massive and lepton
flavors are mixed. A full description of the mass spectrum
and flavor mixing in the framework of three lepton families
requires twelve real parameters: three charged lepton masses
($m_e, m_\mu, m_\tau$), three neutrino masses ($m_1, m_2, m_3$),
three flavor mixing angles ($\theta_x, \theta_y, \theta_z$),
one Dirac-type CP-violating phase ($\delta$) and two Majorana-type
CP-violating phases ($\rho$ and $\sigma$). So far only the masses of
charged leptons have been accurately measured \cite{PDG}. Although we have
achieved some preliminary knowledge on two neutrino mass-squared
differences and three flavor mixing angles
from current neutrino oscillation experiments, much more effort is
needed to determine these parameters precisely. The more challenging
task is to pin down the absolute neutrino mass scale and the
CP-violating phases. Towards reaching these goals, a number of new
neutrino experiments have been proposed \cite{Kayser}.

\vspace{0.25cm}

After sufficient information on neutrino masses and lepton flavor mixing
parameters is experimentally accumulated, a determination of
the textures of lepton mass matrices should become possible. On the
other hand, the textures of charged lepton and neutrino mass matrices
may finally be derived from a fundamental theory of lepton mass
generation, which is unfortunately unknown for the time being.
It is therefore important in phenomenology to investigate how the
textures of lepton mass matrices can link up with the observables of
lepton flavor mixing.

\vspace{0.25cm}

Recently Frampton, Glashow and Marfatia \cite{FGM} have examined the
possibility that a restricted class of lepton mass matrices may
suffice to describe current experimental data. They find seven
acceptable textures of the neutrino mass matrix with two independent
vanishing entries in the flavor basis where the charged lepton mass
matrix is diagonal.

\vspace{0.25cm}

In this paper we carry out a further study of two-zero textures of
the neutrino mass matrix. Our work is different from Ref. \cite{FGM}
in several aspects: (a) we write out the generic constraint equations
for the neutrino mass matrix with two independent vanishing entries, from
which the analytically exact expressions of neutrino mass ratios can be
derived; (b) the formulas to calculate the Majorana-type CP-violating
phases are presented; (c) the relative magnitudes of neutrino masses,
the Majorana phases, the ratio of two neutrino mass-squared differences,
and the effective mass term of the neutrinoless double beta decay are
estimated by taking typical inputs of the flavor mixing angles and
the Dirac-type CP-violating phase; and (d) an order-of-magnitude
illustration is given for seven two-zero textures of the neutrino mass
matrix.

\vspace{0.4cm}

\framebox{\Large\bf 2} ~
In the flavor basis where the charged lepton mass matrix is
diagonal, the neutrino mass matrix can be written as
\begin{equation}
M \; =\; V \left ( \matrix{
m_1 & 0 & 0 \cr
0 & m_2 & 0 \cr
0 & 0 & m_3 \cr} \right ) V^T \; ,
\end{equation}
where $m_i$ (for $i=1,2,3$) denote the real and positive neutrino masses,
and $V$ is the lepton flavor mixing matrix linking the neutrino
mass eigenstates $(\nu_1, \nu_2, \nu_3)$ to the neutrino flavor
eigenstates $(\nu_e, \nu_\mu, \nu_\tau)$ in the chosen basis. A
full description of $V$ needs six real parameters: three mixing angles
and three CP-violating phases. Note that $V$ can always be expressed as
a product of the Dirac-type flavor mixing matrix $U$ (consisting of
three mixing angles and one CP-violating phase) and a diagonal phase
matrix $P$ (consisting of two nontrivial Majorana phases):
$V = UP$. Then we may rewrite $M$ in Eq. (1) as
\begin{equation}
M \; =\; U \left ( \matrix{
\lambda_1 & 0 & 0 \cr
0 & \lambda_2 & 0 \cr
0 & 0 & \lambda_3 \cr} \right ) U^T \; ,
\end{equation}
where two Majorana-type CP-violating phases are included into the complex
neutrino mass eigenvalues $\lambda_i$, and the relation
$|\lambda_i| = m_i$ holds. Without loss of generality, we take
\begin{equation}
\lambda_1 \; =\; m_1 e^{2i\rho} \; , ~~~
\lambda_2 \; =\; m_2 e^{2i\sigma} \; , ~~~
\lambda_3 = m_3 \; .
\end{equation}
In the following we shall show that both the neutrino mass ratios
($m_1/m_3$ and $m_2/m_3$) and the Majorana phases ($\rho$ and
$\sigma$) can be determined, if two independent entries of $M$ vanish.

\vspace{0.25cm}

As $M$ is symmetric, it totally has six independent complex entries.
If two of them vanish, i.e., $M_{ab} = M_{\alpha \beta} =0$,
we obtain the following constraint relations:
\begin{equation}
\sum_{i=1}^{3} \left (U_{ai} U_{bi} \lambda_i \right ) = 0 \; , ~~~~~~
\sum_{i=1}^{3} \left (U_{\alpha i} U_{\beta i} \lambda_i \right ) = 0 \; ,
\end{equation}
where each of the four subscripts runs over
$e$, $\mu$ and $\tau$, but $(\alpha, \beta) \neq (a, b)$.
Solving Eq. (4), we find
\begin{equation}
\frac{\lambda_1}{\lambda_3} \; =\;
\frac{U_{a3} U_{b3} U_{\alpha 2} U_{\beta 2} - U_{a2} U_{b2} U_{\alpha 3}
U_{\beta 3}}{U_{a2} U_{b2} U_{\alpha 1} U_{\beta 1} - U_{a1} U_{b1}
U_{\alpha 2} U_{\beta 2}} \; ,
\end{equation}
and
\begin{equation}
\frac{\lambda_2}{\lambda_3} \; =\;
\frac{U_{a1} U_{b1} U_{\alpha 3} U_{\beta 3} - U_{a3} U_{b3} U_{\alpha 1}
U_{\beta 1}}{U_{a2} U_{b2} U_{\alpha 1} U_{\beta 1} - U_{a1} U_{b1}
U_{\alpha 2} U_{\beta 2}} \; .
\end{equation}
One can observe that the left-hand sides of Eqs. (5) and (6) are
associated with the Majorana-type CP-violating phases, while the
right-hand sides of Eqs. (5) and (6) are associated with the
Dirac-type CP-violating phase hidden in the elements of $U$.
Therefore two Majorana phases must depend upon the Dirac-type
CP-violating phase. This dependence results simply from the texture
zeros of $M$ that we have taken.

\vspace{0.25cm}

Comparing Eq. (5) or Eq. (6) with Eq. (3), we arrive at the expressions of
two neutrino mass ratios as follows:
\begin{eqnarray}
\frac{m_1}{m_3} & = & \left |
\frac{U_{a3} U_{b3} U_{\alpha 2} U_{\beta 2} - U_{a2} U_{b2} U_{\alpha 3}
U_{\beta 3}}{U_{a2} U_{b2} U_{\alpha 1} U_{\beta 1} - U_{a1} U_{b1}
U_{\alpha 2} U_{\beta 2}} \right | \; ,
\nonumber \\ \nonumber \\
\frac{m_2}{m_3} & = & \left |
\frac{U_{a1} U_{b1} U_{\alpha 3} U_{\beta 3} - U_{a3} U_{b3} U_{\alpha 1}
U_{\beta 1}}{U_{a2} U_{b2} U_{\alpha 1} U_{\beta 1} - U_{a1} U_{b1}
U_{\alpha 2} U_{\beta 2}} \right | \; .
\end{eqnarray}
Furthermore, the expressions of two Majorana phases are found to be
\begin{eqnarray}
\rho & = &
\frac{1}{2} \arg \left [ \frac{U_{a3} U_{b3} U_{\alpha 2} U_{\beta 2}
- U_{a2} U_{b2} U_{\alpha 3}
U_{\beta 3}}{U_{a2} U_{b2} U_{\alpha 1} U_{\beta 1} - U_{a1} U_{b1}
U_{\alpha 2} U_{\beta 2}} \right ] \; ,
\nonumber \\ \nonumber \\
\sigma & = & \frac{1}{2} \arg \left [
\frac{U_{a1} U_{b1} U_{\alpha 3} U_{\beta 3} - U_{a3} U_{b3} U_{\alpha 1}
U_{\beta 1}}{U_{a2} U_{b2} U_{\alpha 1} U_{\beta 1} - U_{a1} U_{b1}
U_{\alpha 2} U_{\beta 2}} \right ] \; .
\end{eqnarray}
With the inputs of three flavor mixing angles and the Dirac-type
CP-violating phase, we are able to predict the relative magnitudes of
three neutrino masses and the values of two Majorana phases.
This predictability allows us to examine whether the chosen texture
of $M$ with two independent vanishing entries is empirically acceptable or not.

\vspace{0.25cm}

Indeed the prediction for $m_1/m_2$ and $m_2/m_3$ in a given pattern of
$M$ is required to be compatible with the hierarchy of solar and atmospheric
neutrino mass-squared differences:
\begin{equation}
R_\nu \; \equiv \; \left | \frac{m^2_2 - m^2_1}
{m^2_3 - m^2_2} \right | \; \approx \; \frac{\Delta m^2_{\rm sun}}
{\Delta m^2_{\rm atm}} \; \ll \; 1 \; .
\end{equation}
The magnitude of $R_\nu$ depends upon the explicit solution to the
solar neutrino problem. For the large-angle
Mikheyev-Smirnov-Wolfenstein (MSW) oscillation of solar
neutrinos \cite{MSW}, which is most favored by the present
Super-Kamiokande \cite{SK}
and SNO \cite{SNO} data, we have $R_\nu \sim {\cal O}(10^{-2})$.
Because of $|V_{e3}|^2 = |U_{e3}|^2 \ll 1$ \cite{CHOOZ},
the atmospheric neutrino oscillation is
approximately decoupled from the solar neutrino oscillation.

\vspace{0.25cm}

With the help of Eqs. (7) and (8), one can calculate the effective mass
term of the neutrinoless double beta decay, whose magnitude amounts to
$|M_{ee}|$. The explicit expression of $|M_{ee}|$ reads as follows:
\begin{equation}
| M_{ee} | \; = \; m_3
\left | \frac{m_1}{m_3} U^2_{e1} e^{2i\rho} +
\frac{m_2}{m_3} U^2_{e2} e^{2i\sigma} + U^2_{e3} \right | \; .
\end{equation}
The Heidelberg-Moscow Collaboration has reported
$|M_{ee}| < 0.34$ eV at the $90\%$ confidence
level \cite{Beta}. Useful information on the absolute
mass scale of neutrinos could in principle be extracted from a
more accurate measurement of $|M_{ee}|$ in the future.

\vspace{0.4cm}

\framebox{\Large\bf 3} ~
As already pointed out in Ref. \cite{FGM}, there are totally fifteen logical
possibilities for the texture of $M$ with two independent vanishing entries,
but only seven of them are in accord with current experimental data
and empirical hypotheses. The seven acceptable patterns of $M$ are
listed in Table 1, where all the non-vanishing entries are symbolized
by $\times$'s. To work out the explicit expressions of
$\lambda_1/\lambda_3$ and $\lambda_2/\lambda_3$ in each case, we adopt
the following parametrization for the Dirac-type flavor mixing
matrix $U$:
\begin{equation}
U \; = \; \left ( \matrix{
c_x c_z & s_x c_z & s_z \cr
- c_x s_y s_z - s_x c_y e^{-i\delta} &
- s_x s_y s_z + c_x c_y e^{-i\delta} &
s_y c_z \cr
- c_x c_y s_z + s_x s_y e^{-i\delta} &
- s_x c_y s_z - c_x s_y e^{-i\delta} &
c_y c_z \cr } \right ) \; ,
\end{equation}
where $s_x \equiv \sin\theta_x$, $c_x \equiv \cos\theta_x$, and so on.
The advantage of this phase choice is that the Dirac-type CP-violating
phase $\delta$ does not appear in the effective mass term of the neutrinoless
double beta decay \cite{FX01}. In other words, the latter depends
only upon the Majorana phases $\rho$ and $\sigma$ in our phase convention.
Without loss of generality, three mixing angles
($\theta_x, \theta_y, \theta_z$) can all be arranged to lie in the first
quadrant. Three CP-violating phases ($\delta, \rho, \sigma$) may take
arbitrary values from $-\pi$ to $+\pi$ (or from 0 to $2\pi$).

\vspace{0.25cm}

Now let us calculate $\lambda_1/\lambda_3$ and $\lambda_2/\lambda_3$ for
each pattern of $M$ with the help of Eqs. (5), (6) and (11). The
instructive results for $m_1/m_3$, $m_2/m_3$, $\rho$, $\sigma$,
$R_\nu$ and $|M_{ee}|$ may then be obtained.

\vspace{0.25cm}

\underline{Pattern $\rm A_1:$} ~ $M_{ee} = M_{e\mu} = 0$ (i.e., $a=b=e$;
$\alpha = e$ and $\beta =\mu$). We obtain
\begin{eqnarray}
\frac{\lambda_1}{\lambda_3} & = &
+ \frac{s_z}{c^2_z} \left ( \frac{s_x s_y}{c_x c_y} ~ e^{i\delta}
- s_z \right ) \; ,
\nonumber \\
\frac{\lambda_2}{\lambda_3} & = &
- \frac{s_z}{c^2_z} \left ( \frac{c_x s_y}{s_x c_y} ~ e^{i\delta}
+ s_z \right ) \; .
\end{eqnarray}
As current experimental data favor $\sin^2 2\theta_x \sim {\cal O}(1)$,
$\sin^2 2\theta_y \approx 1$ and
$\sin^2 2\theta_z \leq 0.1$ \cite{SK,SNO,CHOOZ}, one
may make an analytical approximation for the exact result obtained above.
By use of Eqs. (7)--(10), we arrive explicitly at
\begin{eqnarray}
&& \frac{m_1}{m_3} \; \approx \; t_x t_y s_z \; , ~~~
\frac{m_2}{m_3} \; \approx \; \frac{t_y}{t_x} s_z \; ; ~~~
\rho \; \approx \; \frac{\delta}{2} \; , ~~~
\sigma \; \approx \; \frac{\delta}{2} \pm \frac{\pi}{2} \; ;
\nonumber \\
&& R_\nu \; \approx \; \frac{t^2_y}{t^2_x} \left | 1 - t^4_x \right |
s^2_z \; , ~~~
|M_{ee}| \; = \; 0 \;
\end{eqnarray}
to lowest order, where $t_x \equiv \tan\theta_x$ and so on.
Taking the typical inputs $\theta_x = 30^\circ$,
$\theta_y = 40^\circ$, $\theta_z = 5^\circ$ and $\delta = 90^\circ$,
we obtain $m_1/m_3 \approx 0.04$, $m_2/m_3 \approx 0.13$,
$\rho \approx 45^\circ$, and $\sigma \approx 135^\circ$ (or $-45^\circ$).
In addition, we get $R_\nu \approx 0.014$, consistent with our
empirical hypothesis that the solar neutrino deficit is attributed to
the large-angle MSW oscillation. The vanishing of $|M_{ee}|$ implies
that it is in practice impossible to detect the neutrinoless double
beta decay.

\vspace{0.25cm}

\underline{Pattern $\rm A_2$:} ~ $M_{ee} = M_{e\tau} = 0$ (i.e., $a=b=e$;
$\alpha =e$ and $\beta =\tau$). We obtain
\begin{eqnarray}
\frac{\lambda_1}{\lambda_3} & = &
- \frac{s_z}{c^2_z} \left ( \frac{s_x c_y}{c_x s_y} ~ e^{i\delta}
+ s_z \right ) \; ,
\nonumber \\
\frac{\lambda_2}{\lambda_3} & = &
+ \frac{s_z}{c^2_z} \left ( \frac{c_x c_y}{s_x s_y} ~ e^{i\delta}
- s_z \right ) \; .
\end{eqnarray}
In the lowest-order approximation, we explicitly obtain
\begin{eqnarray}
&& \frac{m_1}{m_3} \; \approx \; \frac{t_x}{t_y} s_z \; , ~~~
\frac{m_2}{m_3} \; \approx \; \frac{1}{t_x t_y} s_z \; ; ~~~
\rho \; \approx \; \frac{\delta}{2} \pm \frac{\pi}{2} \; , ~~~
\sigma \; \approx \; \frac{\delta}{2} \; ;
\nonumber \\
&& R_\nu \; \approx \; \frac{1}{t^2_x t^2_y} \left | 1 - t^4_x \right |
s^2_z \; , ~~~
|M_{ee}| \; = \; 0 \; .
\end{eqnarray}
Using the same inputs as above, we get
$m_1/m_3 \approx 0.06$, $m_2/m_3 \approx 0.18$,
$\rho \approx 135^\circ$ (or $-45^\circ$),
$\sigma \approx 45^\circ$, and $R_\nu \approx 0.03$.
We see that the phenomenological consequences of patterns $\rm A_1$
and $\rm A_2$ are nearly the same \cite{FGM}. However,
pattern $\rm A_2$ seems to be more interesting for model
building \cite{FX00}, in particular when the spirit of lepton-quark
similarity is taken into account.

\vspace{0.25cm}

\underline{Pattern $\rm B_1$:} ~ $M_{\mu\mu} = M_{e\tau} = 0$ (i.e., $a=b=\mu$;
$\alpha =e$ and $\beta =\tau$). We obtain
\begin{eqnarray}
\frac{\lambda_1}{\lambda_3} & = &
\frac{s_x c_x s_y \left (2 c^2_y s^2_z - s^2_y c^2_z \right )
- c_y s_z \left ( s^2_x s^2_y e^{+i\delta} + c^2_x c^2_y e^{-i\delta} \right )}
{s_x c_x s_y c^2_y +
\left ( s^2_x - c^2_x \right ) c^3_y s_z e^{i\delta} +
s_x c_x s_y s^2_z \left ( 1 + c^2_y \right ) e^{2i\delta}} ~ e^{2i\delta} \; ,
\nonumber \\
\frac{\lambda_2}{\lambda_3} & = &
\frac{s_x c_x s_y \left (2 c^2_y s^2_z - s^2_y c^2_z \right )
+ c_y s_z \left ( c^2_x s^2_y e^{+i\delta} + s^2_x c^2_y e^{-i\delta} \right )}
{s_x c_x s_y c^2_y +
\left ( s^2_x - c^2_x \right ) c^3_y s_z e^{i\delta} +
s_x c_x s_y s^2_z \left ( 1 + c^2_y \right ) e^{2i\delta}} ~ e^{2i\delta} \; .
\end{eqnarray}
The smallness of $s^2_z$ allows us to make a similar analytical approximation
as before. To lowest order, we find
\begin{eqnarray}
&& \frac{m_1}{m_3} \; \approx \; \frac{m_2}{m_3} \; \approx \; t^2_y \; ; ~~~
\rho \; \approx \; \sigma \; \approx \delta \pm \frac{\pi}{2} \; ;
\nonumber \\
&& R_\nu \; \approx \; \frac{1 + t^2_x}{t_x}
\left | t_{2y} c_\delta \right | s_z \; , ~~~
|M_{ee}| \; \approx \; m_3 t^2_y \; ,
\end{eqnarray}
where $t_{2y} \equiv \tan 2\theta_y$ and $c_\delta \equiv \cos\delta$.
Note that
\begin{equation}
\frac{m_1}{m_3} - \frac{m_2}{m_3} \; \approx \;
\frac{4 s_z c_\delta}{s_{2x} s_{2y}} \; , ~~~
\sigma - \rho \; \approx \; \frac{2 s_z s_\delta}{t^2_y s_{2x} t_{2y}} \;
\end{equation}
in the next-to-leading order approximation, where
$s_\delta \equiv \sin\delta$ and $s_{2x} \equiv \sin 2\theta_x$.
Typically taking $\theta_x = 30^\circ$, $\theta_y = 40^\circ$,
$\theta_z = 5^\circ$ and $\delta = 89^\circ$, we arrive at
$m_1/m_3 \approx m_2/m_3 \approx 0.7$ with a difference of about $0.007$,
$\sigma \approx \rho \approx 179^\circ$ (or $-1^\circ$) with a
difference of about $3^\circ$,
$R_\nu \approx 0.02$, and $|M_{ee}|/m_3 \approx 0.7$.
One can see that $|\delta| \approx 90^\circ$ is required in this texture
of $M$ for plausible inputs of $\theta_y$ and $\theta_z$,
such that $R_\nu$ gets suppressed sufficiently. If pattern
$\rm B_1$ is realistically correct, large CP-violating effects may be
observable in long-baseline neutrino oscillations. It is also worth
mentioning that a typical upper
bound on three nearly degenerate neutrino masses can be extracted from the
Heidelberg-Moscow experiment \cite{Beta}:
$m_1 \approx m_2 \approx 0.7 m_3 \approx |M_{ee}| < 0.34$ eV. This bound
is certainly compatible with the present
direct-mass-search experiments \cite{PDG},
in particular for the electron neutrino.

\vspace{0.25cm}

\underline{Pattern $\rm B_2$:} ~ $M_{\tau\tau} = M_{e\mu} = 0$
(i.e., $a=b=\tau$; $\alpha =e$ and $\beta =\mu$). We obtain
\begin{eqnarray}
\frac{\lambda_1}{\lambda_3} & = &
\frac{s_x c_x c_y \left (2 s^2_y s^2_z - c^2_y c^2_z \right )
+ s_y s_z \left ( s^2_x c^2_y e^{+i\delta} + c^2_x s^2_y e^{-i\delta} \right )}
{s_x c_x s^2_y c_y -
\left ( s^2_x - c^2_x \right ) s^3_y s_z e^{i\delta} +
s_x c_x c_y s^2_z \left ( 1 + s^2_y \right ) e^{2i\delta}} ~ e^{2i\delta} \; ,
\nonumber \\
\frac{\lambda_2}{\lambda_3} & = &
\frac{s_x c_x c_y \left (2 s^2_y s^2_z - c^2_y c^2_z \right )
- s_y s_z \left ( c^2_x c^2_y e^{+i\delta} + s^2_x s^2_y e^{-i\delta} \right )}
{s_x c_x s^2_y c_y -
\left ( s^2_x - c^2_x \right ) s^3_y s_z e^{i\delta} +
s_x c_x c_y s^2_z \left ( 1 + s^2_y \right ) e^{2i\delta}} ~ e^{2i\delta} \; .
\end{eqnarray}
In the lowest-order approximation, we explicitly obtain
\begin{eqnarray}
&& \frac{m_1}{m_3} \; \approx \; \frac{m_2}{m_3} \; \approx \;
\frac{1}{t^2_y} \; ; ~~~
\rho \; \approx \; \sigma \; \approx \delta \pm \frac{\pi}{2} \; ;
\nonumber \\
&& R_\nu \; \approx \; \frac{1 + t^2_x}{t_x}
\left | t_{2y} c_\delta \right | s_z \; , ~~~
|M_{ee}| \; \approx \; \frac{m_3}{t^2_y} \; ,
\end{eqnarray}
together with
\begin{equation}
\frac{m_2}{m_3} - \frac{m_1}{m_3} \; \approx \;
\frac{4 s_z c_\delta}{s_{2x} s_{2y}} \; , ~~~
\sigma - \rho \; \approx \; \frac{2 t^2_y s_z s_\delta}{s_{2x} t_{2y}} \; .
\end{equation}
Using the same inputs as in pattern $\rm B_1$, we get
$m_2/m_3 \approx m_1/m_3 \approx 1.4$ with a difference of about $0.007$,
$\sigma \approx \rho \approx 179^\circ$ (or $-1^\circ$) with a
difference of about $1.4^\circ$,
$R_\nu \approx 0.02$, and $|M_{ee}|/m_3 \approx 1.4$. Because of
$t_y \sim {\cal O}(1)$, the phenomenological consequences of
patterns $\rm B_1$ and $\rm B_2$ are almost the same.

\vspace{0.25cm}

\underline{Pattern $\rm B_3$:} ~ $M_{\mu\mu} = M_{e\mu} = 0$ (i.e., $a=b=\mu$;
$\alpha =e$ and $\beta =\mu$). We obtain
\begin{eqnarray}
\frac{\lambda_1}{\lambda_3} & = &
- \frac{s_y}{c_y} \cdot \frac{s_x s_y - c_x c_y s_z e^{-i\delta}}
{s_x c_y + c_x s_y s_z e^{+i\delta}} ~ e^{2i\delta} \; ,
\nonumber \\
\frac{\lambda_2}{\lambda_3} & = &
- \frac{s_y}{c_y} \cdot \frac{c_x s_y + s_x c_y s_z e^{-i\delta}}
{c_x c_y - s_x s_y s_z e^{+i\delta}} ~ e^{2i\delta} \; .
\end{eqnarray}
The approximate expressions for the neutrino mass ratios, the Majorana phases
and the observables $R_\nu$ and $|M_{ee}|$ turn out to be
\begin{eqnarray}
&& \frac{m_1}{m_3} \; \approx \; \frac{m_2}{m_3} \; \approx \; t^2_y \; ; ~~~
\rho \; \approx \; \sigma \; \approx \delta \pm \frac{\pi}{2} \; ;
\nonumber \\
&& R_\nu \; \approx \; \frac{1 + t^2_x}{t_x} t^2_y
\left | t_{2y} c_\delta \right | s_z \; , ~~~
|M_{ee}| \; \approx \; m_3 t^2_y \; .
\end{eqnarray}
In addition,
\begin{equation}
\frac{m_2}{m_3} - \frac{m_1}{m_3} \; \approx \;
\frac{4 t^2_y s_z c_\delta}{s_{2x} s_{2y}} \; , ~~~
\rho - \sigma \; \approx \; \frac{2 s_z s_\delta}{s_{2x} t_{2y}} \;
\end{equation}
in the next-to-leading order approximation.
Taking the same inputs as in pattern $\rm B_1$, we
find $m_2/m_3 \approx m_1/m_3 \approx 0.7$ with a difference of about $0.005$,
$\rho \approx \sigma \approx 179^\circ$ (or $-1^\circ$) with a
difference of about $2^\circ$,
$R_\nu \approx 0.014$, and $|M_{ee}|/m_3 \approx 0.7$. One can see that
the phenomenological consequences of pattern $\rm B_3$ are essentially
the same as those of pattern $\rm B_1$. This point has been observed in
Ref. \cite{FGM}.

\vspace{0.25cm}

\underline{Pattern $\rm B_4$:} ~ $M_{\tau\tau} = M_{e\tau} = 0$
(i.e., $a=b=\tau$; $\alpha =e$ and $\beta =\tau$). We obtain
\begin{eqnarray}
\frac{\lambda_1}{\lambda_3} & = &
- \frac{c_y}{s_y} \cdot \frac{s_x c_y + c_x s_y s_z e^{-i\delta}}
{s_x s_y - c_x c_y s_z e^{+i\delta}} ~ e^{2i\delta} \; ,
\nonumber \\
\frac{\lambda_2}{\lambda_3} & = &
- \frac{c_y}{s_y} \cdot \frac{c_x c_y - s_x s_y s_z e^{-i\delta}}
{c_x s_y + s_x c_y s_z e^{+i\delta}} ~ e^{2i\delta} \; .
\end{eqnarray}
To lowest order, we get the following approximate results:
\begin{eqnarray}
&& \frac{m_1}{m_3} \; \approx \; \frac{m_2}{m_3} \; \approx \;
\frac{1}{t^2_y} \; ; ~~~
\rho \; \approx \; \sigma \; \approx \delta \pm \frac{\pi}{2} \; ;
\nonumber \\
&& R_\nu \; \approx \; \frac{1 + t^2_x}{t_x t^2_y}
\left | t_{2y} c_\delta \right | s_z \; , ~~~
|M_{ee}| \; \approx \; \frac{m_3}{t^2_y} \; ,
\end{eqnarray}
together with
\begin{equation}
\frac{m_1}{m_3} - \frac{m_2}{m_3} \; \approx \;
\frac{4 s_z c_\delta}{s_{2x} s_{2y} t^2_y} \; , ~~~
\rho - \sigma \; \approx \; \frac{2 s_z s_\delta}{s_{2x} t_{2y}} \; .
\end{equation}
Using the same inputs as in pattern $\rm B_1$, we obtain
$m_1/m_3 \approx m_2/m_3 \approx 1.4$ with a difference of about $0.01$,
$\rho \approx \sigma \approx 179^\circ$ (or $-1^\circ$) with a
difference of about $2^\circ$,
$R_\nu \approx 0.03$, and $|M_{ee}|/m_3 \approx 1.4$.
One can see that the phenomenological consequences of patterns $\rm B_1$,
$\rm B_2$, $\rm B_3$ and $\rm B_4$ are nearly the same. Therefore it is
very difficult, even impossible, to distinguish one of them from the
others in practical experiments. Nevertheless, one of the four textures
might be more favored than the others in model building, when underlying
flavor symmetries responsible for those texture zeros are taken into account.

\vspace{0.25cm}

\underline{Pattern C:} ~ $M_{\mu\mu} = M_{\tau\tau} = 0$ (i.e., $a=b=\mu$
and $\alpha = \beta =\tau$). We obtain
\begin{eqnarray}
\frac{\lambda_1}{\lambda_3} & = &
- \frac{c_x c^2_z}{s_z} \cdot \frac{ c_x \left ( s^2_y - c^2_y \right )
+ 2 s_x s_y c_y s_z e^{i\delta}}
{2 s_x c_x s_y c_y - \left ( s^2_x - c^2_x \right )
\left ( s^2_y - c^2_y \right ) s_z e^{i\delta} + 2 s_x c_x s_y c_y s^2_z
e^{2i\delta}} ~ e^{i\delta} \; ,
\nonumber \\
\frac{\lambda_2}{\lambda_3} & = &
+ \frac{s_x c^2_z}{s_z} \cdot \frac{ s_x \left ( s^2_y - c^2_y \right )
- 2 c_x s_y c_y s_z e^{i\delta}}
{2 s_x c_x s_y c_y - \left ( s^2_x - c^2_x \right )
\left ( s^2_y - c^2_y \right ) s_z e^{i\delta} + 2 s_x c_x s_y c_y s^2_z
e^{2i\delta}} ~ e^{i\delta} \; .
\end{eqnarray}
Assuming $s^2_z \ll 1$ and $t_x \sim t_y \sim {\cal O}(1)$, we may make an
analytical approximation for the exact result in Eq. (28). To lowest order,
we get
\begin{eqnarray}
\frac{m_1}{m_3} & \approx &
\sqrt{1 - \frac{2 c_\delta}{t_x t_{2y} s_z} +
\frac{1}{t^2_x t^2_{2y} s^2_z}} \;\; ,
\nonumber \\
\frac{m_2}{m_3} & \approx &
\sqrt{1 + \frac{2 t_x c_\delta}{t_{2y} s_z} +
\frac{t^2_x}{t^2_{2y} s^2_z}} \;\; ,
\nonumber \\
R_\nu & \approx & \frac{1 + t_x t_y}{t_x t_y}
\left | \frac{2}{t_{2x}} \cdot \frac{1 - t_{2x} t_{2y} s_z c_\delta}
{t_x + 2 t_{2y} s_z c_\delta} \right | \; ,
\nonumber \\
|M_{ee}| & \approx & m_3 \sqrt{1 - \frac{4 c_\delta}{t_{2x} t_{2y} s_z}
+ \frac{4}{t_{2x} t_{2y} s^2_z}} \; \; ;
\end{eqnarray}
as well as
\begin{eqnarray}
\rho & \approx & \delta + \epsilon \pm \frac{\pi}{2}  ~~
{\rm with} ~~ t_\epsilon \; =\; \frac{s_\delta}{t_x t_{2y} s_z - c_\delta} \; ,
\nonumber \\
\sigma & \approx & \delta - \varepsilon \pm \frac{\pi}{2}  ~~
{\rm with} ~~ t_\varepsilon \; =\;
\frac{t_x s_\delta}{t_{2y} s_z + t_x c_\delta} \; .
\end{eqnarray}
One can observe that a small value of $R_\nu$ is possible if and only if
the condition $t_{2x} t_{2y} s_z c_\delta \approx 1$ is satisfied \cite{FGM}.
Some fine tuning of the inputs seems unavoidable in this case. Taking
$\theta_x = \theta_y = 44.8^\circ$, $\theta_z = 5^\circ$ and
$\delta = 90^\circ$ for example, we find $R_\nu \approx 0.03$,
$\rho \approx +5^\circ$ (or $185^\circ$),
$\sigma \approx -5^\circ$ (or $175^\circ$), and
$m_1 \approx m_2 \approx m_3 \approx |M_{ee}|$. If this texture of $M$ is
realistically correct, large CP violation may manifest itself
in neutrino oscillations.

\vspace{0.4cm}

\framebox{\Large\bf 4} ~
As shown above, the seven patterns of $M$ can be classified into three
distinct categories \cite{FGM}: A (with $\rm A_1$ and $\rm A_2$),
B (with $\rm B_1$, $\rm B_2$, $\rm B_3$ and $\rm B_4$), and C.
It is experimentally difficult or impossible to distinguish the
textures of $M$ within each category. However, category A is
experimentally distinguishable from category B or C. To be specific,
let us summarize the main phenomenological consequences of each category:

\vspace{0.25cm}

(1) The neutrino mass spectrum: $m_1 \sim m_2 \ll m_3$ in category A;
$m_1 \sim m_2 \sim m_3$ in category B; and $m_1 \sim m_2 \sim m_3$ in
category C.

\vspace{0.25cm}

(2) The Dirac phase of CP violation: $\delta$ is not constrained in
category A; $|\delta|\approx \pi/2$ in category B (for plausible
inputs of $\theta_y$ and $\theta_z$); and $\delta$ is sensitive
to the values of three mixing angles in category C.

\vspace{0.25cm}

(3) The Majorana phases of CP violation: $|\sigma - \rho | \approx \pi/2$ in
category A; $\sigma \approx \rho$ in category B; and $\sigma \sim \rho$
in category C.

\vspace{0.25cm}

(4) The neutrinoless double beta decay: $|M_{ee}| \approx 0$ in category A;
$|M_{ee}| \sim m_3$ in category B; and $|M_{ee}| \sim m_3$ in category C.

\vspace{0.25cm}

We see that it is not easy to distinguish between category B and
category C, unless the values of flavor mixing angles
$(\theta_x, \theta_y, \theta_z)$ and the ratio of solar and atmospheric
neutrino mass-squared differences ($R_\nu$) can be accurately determined.

\vspace{0.25cm}

It is worth remarking that the ``inverse'' neutrino mass hierarchy
$m_1 \gg m_2 \gg m_3$ cannot be incorporated with three categories
of $M$ discussed above. The reason is simply that such a hierarchy
conflicts with our empirical hypotheses \cite{FGM}, i.e.,
$\Delta m^2_{\rm sun} = |m^2_2 - m^2_1|$ and
$\Delta m^2_{\rm atm} = |m^2_3 - m^2_2|$.
If $m_1 \gg m_2 \gg m_3$ were assumed, one would inevitably be led to
$R_\nu \equiv |m^2_2 - m^2_1|/|m^2_3 - m^2_2| \approx m^2_1/m^2_2 \gg 1$,
contrary to the prerequisite
$R_\nu \approx \Delta m^2_{\rm sun}/\Delta m^2_{\rm atm} \ll 1$ set in
Eq. (9). Therefore we conclude that only the normal hierarchy or
near degeneracy of neutrino masses is allowed for seven two-zero patterns
of the neutrino mass matrix under consideration.

\vspace{0.25cm}

To give an order-of-magnitude illustration of the neutrino mass matrix,
we calculate the elements of $M$ for each pattern by using the formula
\begin{equation}
M_{ab} \; =\; \sum_{i=1}^3 \left (V_{ai} V_{bi} m_i \right )
\; =\; \sum_{i=1}^3 \left (U_{ai} U_{bi} \lambda_i \right ) \;
\end{equation}
and the typical inputs taken before. The rough results are listed in
Table 1. We see that there is no clear hierarchy among the non-vanishing
elements of $M$, unlike the familiar case of quark mass
matrices \cite{FX99}.

\vspace{0.25cm}

Of course, the specific textures of lepton mass matrices cannot be
preserved to all orders or at any energy scales in the unspecified
interactions which generate lepton masses \cite{FGM}.
Nevertheless, those phenomenologically favored textures at low
energy scales may shed light on the underlying flavor symmetries
responsible for the generation of lepton masses at high energy scales.
It is expected that more precise data of neutrino oscillations in the
future could help select the {\it most} favorable pattern of lepton mass
matrices.

\vspace{0.4cm}

The author would like to thank P.H. Frampton for useful discussions during
WIN2002 in Christchurch. He is also grateful to D. Marfatia for helpful
communications via e-mail.
This work was supported in part by National Natural Science
Foundation of China.

\newpage

\begin{table}
\caption{Seven patterns of the neutrino mass matrix $M$ with two independent
vanishing entries, which are in accord with current experimental data and
empirical hypotheses. An order-of-magnitude illustration of $M$ is given
by using typical inputs of $\theta_x$, $\theta_y$, $\theta_z$ and $\delta$,
as explained in the text.}
\begin{center}
\begin{tabular}{ccccc} \hline\hline
Pattern &~~~& Texture of $M$ &~~~& Order of Magnitude \\ \hline
$\rm A_1$
&& $\left ( \matrix{
{\bf 0} & {\bf 0} & \times \cr
{\bf 0} & \times & \times \cr
\times & \times & \times \cr} \right )$
&&
$ \sim m_3 \left ( \matrix{
{\bf 0} ~ & ~ {\bf 0} ~ & ~ .1 \cr
{\bf 0} & .4 & .5 \cr
.1 & .5 & .6 \cr} \right )$
\\ \hline
$\rm A_2$
&& $\left ( \matrix{
{\bf 0} & \times & {\bf 0} \cr
\times & \times & \times \cr
{\bf 0} & \times & \times \cr} \right )$
&&
$\sim m_3 \left ( \matrix{
{\bf 0} ~ & ~ .1 ~ & ~ {\bf 0} \cr
.1 & .4 & .5 \cr
{\bf 0} & .5 & .6 \cr} \right )$
\\ \hline
$\rm B_1$
&& $\left ( \matrix{
\times & \times & {\bf 0} \cr
\times & {\bf 0} & \times \cr
{\bf 0} & \times & \times \cr} \right )$
&&
$\sim m_3 \left ( \matrix{
.7 & .06 & {\bf 0} \cr
.06 & {\bf 0} & .8 \cr
{\bf 0} & .8 & .3 \cr} \right )$
\\ \hline
$\rm B_2$
&& $\left ( \matrix{
\times & {\bf 0} & \times \cr
{\bf 0} & \times & \times \cr
\times & \times & {\bf 0} \cr} \right )$
&&
$\sim m_1 \left ( \matrix{
1. & {\bf 0} & .05 \cr
{\bf 0} & .3 & .8 \cr
.05 & .8 & {\bf 0} \cr} \right )$
\\ \hline
$\rm B_3$
&& $\left ( \matrix{
\times & {\bf 0} & \times \cr
{\bf 0} & {\bf 0} & \times \cr
\times & \times & \times \cr} \right )$
&&
$\sim m_3 \left ( \matrix{
.7 & {\bf 0} & .07 \cr
{\bf 0} & {\bf 0} & .8 \cr
.07 & .8 & .3 \cr} \right )$
\\ \hline
$\rm B_4$
&& $\left ( \matrix{
\times & \times & {\bf 0} \cr
\times & \times & \times \cr
{\bf 0} & \times & {\bf 0} \cr} \right )$
&&
$\sim m_1 \left ( \matrix{
1. & .04 & {\bf 0} \cr
.04 & .3 & .8 \cr
{\bf 0} & .8 & {\bf 0} \cr} \right )$
\\ \hline
$\rm C$
&& $\left ( \matrix{
\times & \times & \times \cr
\times & {\bf 0} & \times \cr
\times & \times & {\bf 0} \cr} \right )$
&&
$\sim m_3 \left ( \matrix{
1. & .06 & .2 \cr
.06 & {\bf 0} & 1. \cr
.2 & 1. & {\bf 0} \cr} \right )$
\\ \hline\hline
\end{tabular}
\end{center}
\end{table}
\normalsize

\end{document}